\documentclass[11pt]{article}%
\usepackage{fleqn,cospar}
\usepackage{url}
\usepackage{graphicx}
\usepackage[figuresright]{rotating}
\usepackage{amsmath}
\usepackage{amsfonts}
\usepackage{amssymb}%
\begin{document}

\title{Why helicity injection causes coronal flux tubes to develop an axially
invariant cross-section}
\author{P. M. Bellan\address{MC 128-95, Caltech, Pasadena CA 91125, USA}}
\maketitle

\begin{abstract}
It is shown that electric current flowing along an axially non-uniform
magnetic flux tube produces an associated non-linear, non-conservative axial
MHD force which pumps plasma from regions where the flux tube diameter is
small to regions where it is large. In particular, this force will ingest
plasma into the ends\ of a fat, initially potential flux tube and then pump
the ingested plasma towards the middle bulge, thereby causing mass
accumulation at the bulge. \ The ingested plasma convects frozen-in toroidal
magnetic flux which accumulates at the middle as well. Flux accumulation at
the bulge has the remarkable consequence of causing the bulge to diminish so
that the flux tube becomes axially uniform as observed in coronal loops.
Stagnation of the convergent plasma flow at the middle heats the plasma. A
small number of tail particles bouncing synchronously between approaching
fluid elements can be Fermi-accelerated to very high energies. Since driving a
current along a flux tube is tantamount to helicity injection into the flux
tube, this mass ingestion, heating, and straightening should be ubiquitous to
helicity injection processes.

\end{abstract}

\section{\bigskip Introduction}

Remarkably detailed images of coronal\ loops\ provided by the TRACE spacecraft
[Aschwanden et al (2000)] indicate that these loops have cross-sectional area
varying by only 10-20\%\ over their entire length; this surprising behavior
cannot be explained by potential or force-free field models [Klimchuk (2000].
We present here a model explaining this behavior. Since the loops are
typically twisted by a fraction of a turn over their length, they contain
magnetic helicity and so our model should be intrinsic to helicity injection processes.

A coronal loop will be represented here by an axisymmetric flux tube with
straight axis (cf. top, Fig. 1) and cylindrical coordinates $(r,\phi,z)$ will
be used. The loop is characterized by an initially potential poloidal flux
function
\begin{equation}
\psi(r,z)=\int_{0}^{r}B_{z}(r^{\prime},z)2\pi r^{\prime}dr^{\prime}
\label{pflux}%
\end{equation}
with associated poloidal field
\begin{equation}
\mathbf{B}_{pol}=\frac{1}{2\pi}\nabla\psi\times\nabla\phi. \label{Bpol}%
\end{equation}
Axial non-uniformity corresponds to $\psi$ being $z$-dependent and bulging
corresponds to $\psi^{-1}\partial^{2}\psi/\partial z^{2}>0.$ Similarly, the
poloidal current is written as
\begin{equation}
I(r,z)=\int_{0}^{r}J_{z}(r^{\prime},z)2\pi r^{\prime}dr^{\prime}%
\ \label{current}%
\end{equation}
with associated poloidal current density%
\begin{equation}
\mathbf{J}_{pol}=\frac{1}{2\pi}\nabla I\times\nabla\phi. \label{Jpol}%
\end{equation}

\bigskip

\qquad We show that establishing a steady-state current $I$ involves three
sequential stages having distinct physics. The first stage, \textquotedblleft
ramp-up\textquotedblright, has physics akin to a linear Alfv\'{e}n wave, but
it is assumed that the ramp-up rate is sufficiently slow that the effective
Alfv\'{e}n wavelength is infinite. This means that retarded time effects due
to wave propagation issues are negligible and the current ramps up everywhere
simultaneously as in an ordinary electrical circuit. The second stage,
\textquotedblleft axial flow\textquotedblright, has $\partial I/\partial t=0$,
but is not in MHD\ equilibrium because unbalanced, non-conservative
$\mathbf{J\times B}$ forces exist which drive plasma flows. The third stage,
\textquotedblleft stagnation\textquotedblright, involves convection of
magnetic flux by the flows, plasma heating as a result of flow stagnation, and
straightening of the $\psi$ profile until MHD\ equilibrium is established.

\bigskip

\section{First stage (Ramp-up)}

\bigskip

\qquad We represent the current ramp-up by the time-dependence%
\begin{equation}
I(t)=I_{0}\frac{1+\tanh(t/\tau)}{2} \label{currenttimedep}%
\end{equation}
where the ramp-up time is assumed to be much longer than the time it takes for
an Alfv\'{e}n wave to propagate the length $h$ of the flux tube, i.e.,
$\tau>>h/v_{A}.$ Alfv\'{e}n wave propagation effects are therefore unimportant
in which case the system behaves like an electric circuit. From Ampere's law
the toroidal magnetic field is%
\begin{equation}
B_{\phi}(r,z,t)\ =\frac{\mu_{0}I(r,z,t)}{2\pi r}. \label{Btor}%
\end{equation}
The toroidal component of Faraday's law is
\begin{equation}
\frac{\partial E_{r}}{\partial z}-\frac{\partial E_{z}}{\partial r}%
=-\frac{\partial B_{\phi}}{\partial t}. \label{FaradayTor}%
\end{equation}
We note that $B_{\phi}$ has minimal $z$ dependence and that Ohm's law implies
$E_{z}\simeq0$. Thus, integration of Eq.(\ref{FaradayTor}) with respect to $z$
gives%
\begin{equation}
E_{r}\simeq-\ \frac{\mu z\ }{2\pi r}\frac{\partial I}{\partial t} \label{Er}%
\end{equation}
where on the basis of symmetry the location $z=0$ is set to be at the axial
midpoint of the flux tube. Since there is no axial force in this stage,
$U_{z}\ $remains zero and so the radial component of the ideal Ohm's law gives%
\begin{equation}
U_{\phi}=-\frac{E_{r}}{B_{z}}\simeq\frac{\mu z\ }{2\pi rB_{z}}\frac{\partial
I}{\partial t}, \label{Uphi}%
\end{equation}
showing that $U_{\phi}\ $is finite only when $I$ is changing. The change in
$U_{\phi}$ (toroidal acceleration)\ implies the existence of a radial current
determined from the toroidal component of the equation of motion%
\begin{equation}
\rho\frac{\partial U_{\phi}}{\partial t}=-J_{r}B_{z}. \label{tormotion}%
\end{equation}
This current is just the polarization current [Chen(1984), p.40]%
\begin{equation}
J_{r}=-\frac{\rho}{B_{z}}\frac{\partial U_{\phi}}{\partial t}=\frac{\rho
}{B_{z}^{2}}\frac{\partial E_{r}}{\partial t}\ =-\frac{\mu z\rho\ }{2\pi
rB_{z}^{2}}\frac{\partial^{2}I}{\partial t^{2}}. \label{Jr}%
\end{equation}
The transient toroidal velocity given by Eq.(\ref{Uphi}) results in an
azimuthal displacement of the plasma,
\begin{equation}
r\Delta\phi=\int_{0}^{t}U_{\phi}dt=\frac{\mu_{0}zI\ }{2\pi rB_{z}}%
=\frac{zB_{\phi}}{B_{z}} \label{azdisplacement}%
\end{equation}
showing that the plasma motion in this stage follows the twisting of the
magnetic field (the field line can be thought of as being frozen to the plasma
so that when the field line twists, so does the plasma). Thus $J_{r}$ is
finite only when $I$ is changing and $J_{r}$ is first negative and then
positive, corresponding to toroidal acceleration followed by toroidal
deceleration. The $r$ direction here is really a proxy for the $\nabla\psi$
direction, since the polarization current is in the $\nabla\psi$ direction
(because the flux tube is long and slender, the $r$ direction is nearly the
same as the $\nabla\psi$ direction).

\qquad Once $I$ has been established, both $U_{\phi}$ and the polarization
current $J_{r}$ remain zero. The poloidal flux function $\psi$ at this stage
has not changed from its original value --- all that has happened is that a
toroidal field has been added so that the total magnetic field is now%
\begin{equation}
\mathbf{B=}\frac{1}{2\pi}\left(  \nabla\psi\times\nabla\phi+\mu_{0}I\nabla
\phi\right)  . \label{Btot}%
\end{equation}
The original potential flux tube has become twisted as shown in the bottom
sketch of Fig. 1. Since $\psi$ is unchanged from its potential value, the
bottom sketch in Fig.1 has the same poloidal profile (envelope)\ as the top
sketch. This means that
\begin{equation}
\mu_{0}J_{\phi}=r\nabla\phi\cdot\nabla\times\frac{\left(  \nabla\psi
\times\nabla\phi\right)  }{2\pi}=-\frac{r}{2\pi}\nabla\cdot\left(  \frac
{1}{r^{2}}\nabla\psi\right)  =0. \label{Jtor}%
\end{equation}
%

\begin{figure}
[ptb]
\begin{center}
\includegraphics[
trim=0.000000in 0.514388in 0.514971in 0.000000in,
height=4.3539in,
width=5.9785in
]%
{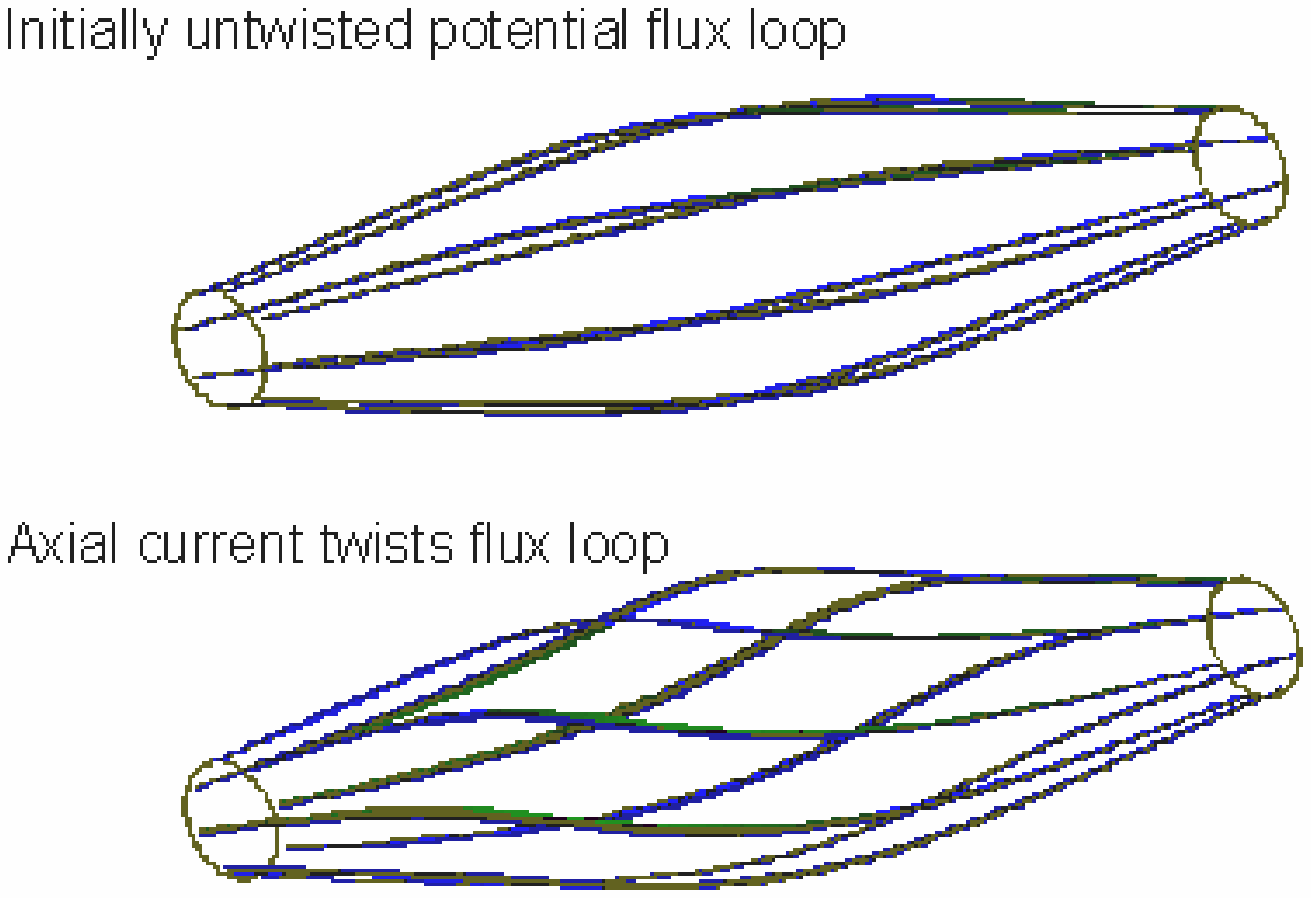}%
\caption{Top:\ Initially potential flux tube. Bottom:\ Flux tube with axial
current has same poloidal profile, but is twisted. Cylindrical geometry
$(r,\phi,z)$ is used; $z=0$ is the flux tube axial midpoint and the flux tube
ends are at $z=\pm h$.}%
\end{center}
\end{figure}
\bigskip

\section{Second stage (flow acceleration)}

\bigskip

\qquad The magnetic force is
\begin{align}
\mathbf{J\times B}  &  \mathbf{=}\left(  \mathbf{J}_{pol}+\mathbf{J}%
_{tor}\right)  \mathbf{\times}\left(  \mathbf{B}_{pol}+\mathbf{B}_{tor}\right)
\nonumber\\
&  =\mathbf{J}_{pol}\mathbf{\times B}_{pol}+\mathbf{J}_{pol}\times
\mathbf{B}_{tor}+\mathbf{J}_{tor}\times\mathbf{B}_{pol}. \label{JxB}%
\end{align}
The term $\mathbf{J}_{pol}\mathbf{\times B}_{pol}$ provides acceleration in
the toroidal direction and involves the component of $\mathbf{J}_{pol}$ which
is perpendicular to $\mathbf{B}_{pol}.$ However, in the previous section we
showed that the component of $\mathbf{J}_{pol}$ which is perpendicular to
$\mathbf{B}_{pol}$ is the polarization current and that this current scales as
$\partial^{2}I/\partial t^{2}.$ Thus, when $I$ is in steady state there is no
polarization current and no toroidal acceleration, and so $U_{\phi}$ remains
zero. It was also shown that $\mathbf{J}_{tor}=J_{\phi}\hat{\phi}$ is zero so
long as the poloidal flux surfaces are unperturbed from their initial
potential shape. We assume in this second stage that (i)\ the current is
constant in which case $\mathbf{J}_{pol}\mathbf{\times B}_{pol}=0$ and
$U_{\phi}=0$ $\ $and (ii)\ the poloidal flux surfaces are unperturbed from
their initial potential shape in which case $\mathbf{J}_{tor}=0.$ Thus, during
this second stage the magnetic force reduces to
\begin{equation}
\mathbf{J\times B=J}_{pol}\times\mathbf{B}_{tor}=\frac{1}{2\pi}\left(
\nabla\psi\times\nabla\phi\right)  \times\frac{\mu_{0}I}{2\pi}\nabla
\phi=-\frac{\mu_{0}}{8\pi^{2}r^{2}}\nabla I^{2}. \label{JxB1}%
\end{equation}
The above force is non-conservative (i.e., $\nabla\times\left(
\mathbf{J\times B}\right)  $ is non-zero) and so cannot be balanced by a
pressure gradient since a pressure gradient is conservative (i.e.,
$\nabla\times\nabla P=0$). Thus, it is not possible for equilibrium to be
attained in this stage. The only way for an equilibrium to be achieved is for
the poloidal profile of the magnetic field to change, which is what happens in
the third stage, to be discussed later.

The fact that $\mathbf{J}_{pol}\mathbf{\times B}_{pol}=0$ means that $\left(
\nabla I\times\nabla\phi\right)  \times\left(  \nabla\psi\times\nabla
\phi\right)  =0$ which in turn implies that $\nabla I$ is parallel to
$\nabla\psi$ and so $I$ must be a function of $\psi,$ i.e., $I=I(\psi).$ Thus,
the poloidal current flows along the poloidal flux surfaces. This is
consistent with the well-known Hamiltonian dynamics concept that, because of
conservation of canonical angular momentum, particles in a toroidally
symmetric geometry cannot make an excursion exceeding a poloidal Larmor radius
from a poloidal flux surface [e.g., see p.207-208 of Bellan(2000)]. In other
words, Hamiltonian mechanics forbids the existence of steady current in the
direction normal to a poloidal flux surface.

As sketched in Fig. 1, the poloidal flux function is bulged near $z=0$,
corresponding to a weaker magnetic field near $z=0$ than at the ends $z=\pm
h.$ This would be the situation if the source currents for the poloidal field
were located external to the flux tube and so the middle of the flux tube
would be further from the source currents than the ends. Since $I=I(\psi)$,
the current channel would also be bulged.

Equation (\ref{JxB1}) implies that the $z$ component of the equation of motion
is
\begin{equation}
\rho\frac{dU_{z}}{dt}=\left(  \mathbf{J\times B}\right)  _{z}=-\frac{1}%
{8\pi^{2}r^{2}}\frac{\partial I^{2}}{\partial z}=-\ \frac{\partial\ }{\partial
z}\left(  \frac{B_{\phi}^{2}}{2\mu_{0}}\right)  . \label{axialaccn}%
\end{equation}
This means that there is a force accelerating plasma from regions where
$B_{\phi}^{2}$ is strong to regions where $B_{\phi}^{2}$ is weak. Since
$I=I(\psi)$ and since $\psi$ is bulged in the middle, $B_{\phi}$ must be
stronger near $z=\pm h$ where the current channel diameter is small than at
$z=0$ where the current channel diameter is large. There consequently must be
an acceleration of plasma from both ends (i.e., $z=\pm h$) towards the middle
(i.e., $z=0$) as shown in Fig. 2. The convergent axial pumping is similar to
the \textquotedblleft sweeping magnetic twist mechanism\textquotedblright%
\ discussed by Uchida and Shibata (1988), but it should not be considered a
wave because it involves actual convection of material and not propagation of
energy through a material. We note in passing that there could be a few
exceptional particles collisionally bouncing back and forth between the
approaching fluid elements. These exceptional particles would be accelerated
to very high energy by the Fermi acceleration process, and so one would expect
to see a tail of energetic particles develop in the vicinity of $z=0.$ The
Fermi process would thus predict that the most energetic particles would be
located around the top of an arched coronal loop and such is indeed what is
observed (Feldman, 2002).%

\begin{figure}
[ptb]
\begin{center}
\includegraphics[
trim=0.514316in 1.028775in 0.514971in 1.029266in,
height=2.8315in,
width=5.4707in
]%
{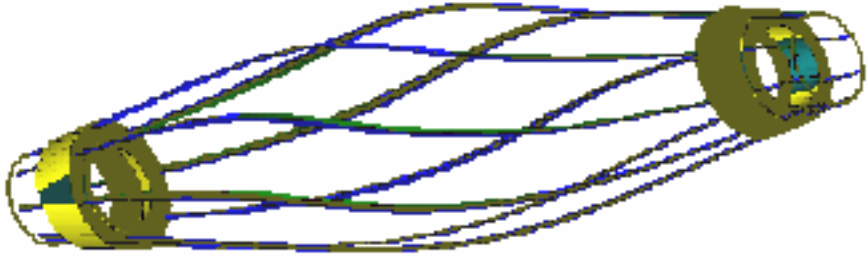}%
\caption{Toroidal plasma fluid elements are accelerated from $z\pm h$ to $z=0$
by force $F_{z}=-\partial\left(  B_{\phi}^{2}/2\mu_{0}\right)  /\partial z.$
These representative fluid elements move towards each other, but do not
rotate.}%
\end{center}
\end{figure}
\bigskip

\section{Third stage (stagnation, heating, and straightening)}

\bigskip

\qquad The flows from both ends converge at the middle and must come to a halt
at $z=0.$ Convergence of flows means that $\nabla\cdot\mathbf{U\ }$is
negative. This has important implications for the magnetic field as can be
seen by considering the induction equation toroidal component,
\begin{equation}
\frac{\partial B_{\phi}}{\partial t}=r\ \mathbf{B}_{pol}\mathbf{\cdot}%
\nabla\left(  \frac{U_{\phi}}{r}\right)  -r\mathbf{U}_{pol}\mathbf{\cdot
}\nabla\left(  \frac{B_{\phi}}{r}\right)  -B_{\phi}\nabla\cdot\mathbf{U}%
_{pol}. \label{torBphi}%
\end{equation}
We have shown that (i) $U_{\phi}=0,$ (ii) $\mathbf{U}_{pol}\rightarrow0$ at
the stagnation layer, and (iii) $\nabla\cdot\mathbf{U}_{pol}$ is negative.
Thus, in the vicinity of the stagnation layer%
\begin{equation}
\frac{\partial B_{\phi}}{\partial t}=-B_{\phi}\nabla\cdot\mathbf{U}%
_{pol}\ \label{Bphistag}%
\end{equation}
showing that $B_{\phi}$ must increase at the stagnation layer (increase of
magnetic field at regions of local flow convergence has been discussed in a
more general context by Polygiannakis and Moussas (1999)).

The continuity equation in the vicinity of the stagnation layer gives
$\nabla\cdot\mathbf{U}_{pol}=-\rho^{-1}\partial\rho/\partial t$ and so in the
vicinity of the stagnation layer%
\begin{equation}
\frac{1}{B_{\phi}}\frac{\partial B_{\phi}}{\partial t}=\frac{1}{\rho}%
\frac{\partial\rho}{\partial t}. \label{normBhi}%
\end{equation}
Thus, in the vicinity of the stagnation layer $B_{\phi}$ increases in
proportion to the accumulation of mass. Since $I$ is constant during this
stage, if $B_{\phi}$ increases, the radius of the current channel must
decrease so as to maintain $\mu_{0}I=2\pi rB_{\phi}$ constant. However, since
$I=I(\psi)$, if the radius of the current channel decreases, then the radius
of $\psi$ must also decrease, thereby reducing the bulging. Ultimately, the
bulge becomes vanishingly small as more and more mass accumulates at the
stagnation point and eventually a plasma-loaded, axially uniform flux tube results.

\qquad Since stagnation involves conversion of flow velocity into thermal
energy, the plasma will be heated at the stagnation layer. The resulting
plasma temperature can be estimated from the details of the stagnation
process. The axial magnetic force in Eq.(\ref{axialaccn}) involves the
gradient of the toroidal field energy density and so the kinetic energy gained
by the plasma will be of the order of the change in toroidal field energy
density $B_{\phi}^{2}/2\mu_{0}$ between the ends ($z=\pm h$) and the midpoint
$z=0.$ Because $B_{\phi}^{2}/2\mu_{0}$ is larger at the ends, the plasma flow
kinetic energy is $\rho U_{z}^{2}/2\simeq$ $\left[  B_{\phi}^{2}/2\mu
_{0}\right]  _{z=\pm h}$ and at the stagnation layer $z=0$ this flow energy is
converted into heat so $\rho U_{z}^{2}=\left[  B_{\phi}^{2}/\mu_{0}\right]
_{z=\pm h}$ $\rightarrow nm_{e}v_{Te}^{2}+nm_{i}v_{Ti}^{2}.\ $

\qquad Since the poloidal field is much stronger than the toroidal field and
is approximately $B_{z},$ this means that the flow stagnation causes the
plasma to develop a state where
\begin{equation}
\beta=\frac{2\mu_{0}(nm_{e}v_{Te}^{2}+nm_{i}v_{Ti}^{2})}{B_{z}^{2}}%
=2\frac{B_{\phi}^{2}}{B_{z}^{2}}=2\left(  \frac{\mu_{0}I}{2\pi aB_{z}}\right)
^{2}\ =\left(  \frac{\mu_{0}I}{\psi}\right)  ^{2}\frac{a^{2}}{2}%
,\ \label{beta}%
\end{equation}
where $a$ is the radius of the current channel. However, $\mu_{0}I=\int
d\mathbf{s}\cdot\nabla\times\mathbf{B}$ and $\psi=\int d\mathbf{s\cdot B}$
where the surface integral is over the cross-sectional area of the flux tube.
If we define the ratio of poloidal current to poloidal flux as%
\begin{equation}
\alpha=\mu_{0}I/\psi, \label{defalpha}%
\end{equation}
the $\beta$ predicted from flow stagnation is%
\begin{equation}
\beta=\ \alpha^{2}a^{2}/2. \label{Beta1}%
\end{equation}

\qquad Thus, when a current is made to flow along an initially bulging
current-free current channel, the current channel will twist up (helicity
injection), plasma will be ingested from both ends, accelerated toward the
middle where it accumulates and heats up at a stagnation layer. The
equilibrium will become straight (filamentary) and have $\beta=\alpha^{2}%
a^{2}/2$ where $\alpha=\mu_{0}I/\psi$ and $a$ is the radius of the current
channel. To an outside observer the current channel will look field-aligned
since the current is axially uniform and appears to be embedded in an axially
uniform axial magnetic field. However, oblivious to the outside observer,
within the current channel there is a Bennett pinch-like radial force balance
between plasma pressure pushing out and $J_{z}B_{\phi}$ magnetic force pushing in.

\qquad Although the predicted $\beta$ is typically very small, its effect is
crucial. To see this, consider that in equilibrium $\mathbf{J\times B}=\nabla
P$ so that $\mathbf{B\cdot}\nabla P=(2\pi)^{-1}\left(  \nabla\psi\times
\nabla\phi+\mu_{0}I\nabla\phi\right)  \mathbf{\cdot}\nabla P=0$ and so $P$
must be a function of $\psi,$ i.e., $P=P(\psi).$ Defining $\psi_{0}$ as the
flux on the flux surface where $P$ vanishes, we can write $P(r,z)=\ (1-\psi
(r,z)/\psi_{0})P_{0}$ where $P_{0}$ is the on-axis pressure (i.e., where
$\psi=0$). We can also write $\mu_{0}I(r,z)=$ $\alpha\psi(r,z)$ and so
$\mathbf{J\times B}=\nabla P$ can be written in Grad-Shafranov form as
\begin{equation}
r\frac{\partial}{\partial r}\left(  \frac{1}{r}\frac{\partial\psi}{\partial
r}\right)  +\frac{\partial^{2}\psi}{\partial z^{2}}+\alpha^{2}\psi=4\pi^{2}%
\mu_{0}r^{2}\frac{P_{0}}{\psi_{0}}. \label{GradShafranov}%
\end{equation}
If $\alpha^{2}=4\pi^{2}\mu_{0}a^{2}P_{0}/\psi_{0}^{2}$ where $a$ is the flux
tube radius at $z=0,$ then the only solution to Eq.(\ref{GradShafranov})
satisfying the specified boundary condition that $P$ vanishes when $\psi
=\psi_{0}$ is the particular solution $\psi(r,z)=\psi_{0}r^{2}/a^{2}$. This
means that the flux tube\textit{\ must} be axially uniform when $\alpha
^{2}=4\pi^{2}\mu_{0}a^{2}P_{0}/\psi_{0}^{2}.$ Defining $B_{0}=\psi_{0}/\pi
a^{2}$ as the axial field at $z=0,$ it is seen that this condition for axial
uniformity corresponds to $\alpha^{2}=4\pi^{2}\mu_{0}a^{2}P_{0}/B_{0}^{2}%
\pi^{2}a^{4}$ or $\alpha^{2}a^{2}/2=\beta$ where $\beta=2\mu_{0}P_{0}%
/B_{0}^{2}.$ This equilibrium has $J_{\phi}=0,$ so all confinement is provided
by the Bennett pinch force $\sim J_{z}B_{\phi}.$ The current is purely in the
$z$ direction, but the magnetic field is helical.

The situation of small but finite $\beta$ is substantially different from the
case of zero $\beta$ because the system is constrained to be axially uniform
if and only if $\beta=\alpha^{2}a^{2}/2.$ The arguments presented in the
discussion of Eqs.(\ref{beta}-\ref{Beta1}) show that the MHD\ dynamical
pumping tends to produce precisely the situation where $\beta=\alpha^{2}%
a^{2}/2,$ and so it is predicted that MHD\ dynamical pumping will always cause
configurations to tend towards being axially uniform\ (i.e., filamentary),
hot, and dense, and with $\beta=\alpha^{2}a^{2}/2.$

The definition of $\alpha$ in Eq.(\ref{defalpha}) is closely related to that
used for force-free fields. However, there is an important difference because
while Eq.(\ref{defalpha}) corresponds to having $\mu_{o}J_{z}=\alpha B_{z}$,
Eq.(\ref{defalpha}) makes no statement about any relationship between
$J_{\phi}$ and $B_{\phi}$. The finite $\beta$ equilibrium discussed in the
previous two paragraphs is\textit{\ not} force-free and involves the radial
force balance $J_{z}B_{\phi}=-\partial P/\partial r\ $with $J_{\phi}%
=0;\ \,\ $this differs from the force-free radial equation $J_{\phi}%
B_{z}-J_{z}B_{\phi}=0$ with $\mu_{o}J_{\phi}=\alpha B_{\phi},$ $\mu_{o}%
J_{z}=\alpha B_{z}.$ It is worth noting that the determination of $\alpha$
made from vector magnetographs (e.g., Pevtsov et al., 1997) effectively use
the definition $\alpha=\mu_{o}J_{z}/B_{z}$ which is equivalent to
Eq.(\ref{defalpha}); these measurements do not provide information on either
$J_{\phi}$ or $B_{\phi}$ and so do not provide any information on the value of
$\mu_{o}J_{\phi}/B_{\phi}.$ Thus there is only one definition for $\alpha,$
but its application is different for force-free situations compared to
finite-$\beta$ situations: for force-free situations $\alpha$ gives the ratio
of current to flux for \textit{both} toroidal and poloidal directions whereas
for the finite $\beta$ situation, $\alpha$ refers \textit{only} to the the
ratio of poloidal current to poloidal field.

\qquad The prediction that $\beta=\alpha^{2}a^{2}/2$ can be compared with the
actual observed values of $\beta$ in TRACE\ flux loops. To calculate the
predicted $\beta,$ we use the nominal \textit{measured} flux loop radius
$a=1.6\times10^{6}$ m from Aschwanden et al.(2000) and the nominal
\textit{measured} active region $\alpha=2\times10^{-8}$ m$^{-1}$ from Fig. 4
of Pevtsov et al. (1997). These parameters give a nominal $\beta
_{predicted}=\alpha^{2}a^{2}/2=\allowbreak5\times10^{-4}.$ The observed value
$\beta_{observed}$ is calculated using a nominal measured density
$n=10^{15}\,\ $m$^{-3}\ $ and a nominal measured temperature $10^{6}$ K
[Aschwanden et al. (2000)]. In addition a nominal axial magnetic field
$B_{z}=1.5\times10^{-2}\,$\ T is assumed based on the argument that because
the flux tube is axially uniform, its axial field must also be axially uniform
and so will have the same value as the nominal $B_{z}=1.5\times10^{-2}\,$\ at
the surface of an active region. These parameters give $\beta_{observed}%
=2\mu_{0}n\kappa T/B_{z}^{2}=4\times10^{-4}$ which is very close to
$\beta_{predicted}$. If the model were wrong, one would expect no relationship
between the predicted and observed $\beta$'s, i.e., one would expect a
discrepancy of many orders of magnitude between the predicted and the observed
$\beta$'s.

\qquad This model also has implications regarding the brightening typically
observed when the axis of a coronal loop starts to writhe and the loop
develops a kink instability (sigmoid). Since kink instability occurs when
$\alpha h\sim1$ and for a long thin flux tube $a<<h$, this model predicts that
$\beta=\alpha^{2}a^{2}/2<<$ $\alpha^{2}h^{2}/2$ will still be small even if
$\alpha$ is increased to the point where $\alpha h\sim1$ and kink instability
occurs. However, $\beta$ will increase as $\alpha$ increases and so this model
predicts that the loop should brighten in proportion to the writhing of its
axis (i.e., in proportion to $\alpha$ as $\alpha h$ approaches unity).

Finally, we note that Feldman (2002) has recently used purely observational
evidence to argue that electric currents with geometry similar to what is
discussed here are the means by which the Sun and similar stars produce their
coronal activity.

Acknowledgment:\ Supported by USDOE\ Grant DE-FG03-97ER54438.

\bigskip

\noindent\textbf{REFERENCES}

Aschwanden, M. J., Nightingale, R. W., and Alexander, D. \ 2000, ApJ 541, 1059

Bellan, P. M., Spheromaks (Imperial College Press, 2000, London)

Chen, F. F., Introduction to Plasma Physics and Controlled Fusion (Plenum
1984, New York)

Feldman, U. 2002, Physica Scripta 65, 1985

Klimchuk, J. A. 2000, Solar Physics 193\textbf{,} 53

Pevtsov, A. A., Canfield, R. C., and McClymont, A. N. 1997, ApJ 481, 973

Polygiannakis, J. M. and Moussas, X. 1999, Plasma Phys. Control. Fusion 41, 967

Uchida Y. and Shibata, K. 1988, Solar Phys. 116, 291

\bigskip

\end{document}